\documentclass[11pt]{article}
\newcommand{\be}{\begin{eqnarray}}\newcommand{\beq}{\begin{equation}}
\newcommand{\ee}{\end{eqnarray}}\newcommand{\eeq}{\end{equation}}

\newcommand{\la}{\lambda}\newcommand{\De}{\Delta}
\input epsf 
\usepackage{graphicx} 
\usepackage{amssymb}
\usepackage{wasysym}
\usepackage[usenames, dvipsnames]{color}

\title{
Free energy of formation of a crystal nucleus in incongruent solidification: Implication for
modeling the crystallization of aqueous nitric acid droplets in type 1 polar stratospheric clouds
}

\author{Yuri S. Djikaev\thanks{Corresponding author. E-mail: idjikaev@buffalo.edu}\hspace{0.2cm} 
and \hspace{0.1cm} Eli Ruckenstein$^{}$\thanks{
E-mail: feaeliru@buffalo.edu }\hspace{0.2cm} \\ 
\\ Department of Chemical and Biological  Engineering, SUNY at Buffalo, \\ 
Buffalo, New York  14260 }
\date{ \hfill }
\renewcommand{\baselinestretch}{2}
\begin{document}
\renewcommand{\baselinestretch}{1}
\maketitle
\renewcommand{\baselinestretch}{1}
{\bf Abstract.} 
{\small 

Using the formalism of classical thermodynamics in the framework of  the classical nucleation
theory, we derive an expression for the reversible work $W_*$ of  formation of a binary crystal
nucleus in a liquid binary solution of non-stoichiometric composition  (incongruent crystallization). Applied 
to the crystallization of aqueous nitric acid droplets, the new expression  more adequately takes  account of
the effects of nitric acid vapor compared to the conventional expression of MacKenzie, Kulmala, Laaksonen,
and Vesala (MKLV) [J.Geophys.Res. {\bf 102}, 19729 (1997)].  The predictions of both MKLV and modified
expressions for the average  liquid-solid interfacial tension $\sigma^{\mbox{\tiny ls}}$ of nitric acid
dihydrate (NAD) crystals are compared by using existing experimental data on the incongruent crystallization
of aqueous nitric acid droplets of composition relevant to polar stratospheric clouds (PSCs). The predictions
for  $\sigma^{\mbox{\tiny ls}}$ based on  the MKLV expression are higher by about 5\% compared to predictions
based on our modified expression. This results in similar differences between the predictions of both
expressions for the solid-vapor interfacial tension $\sigma^{\mbox{\tiny sv}}$ of NAD crystal nuclei.  The
latter can be obtained by using  the method based on  the analysis of  experimental data  on crystal
nucleation rates in aqueous nitric acid droplets; it exploits the dominance of the surface-stimulated mode of
crystal nucleation in small droplets   and its negligibility in large ones. Applying that method to existing
experimental data,  our expression for the free energy of formation provides an  estimate for
$\sigma^{\mbox{\tiny sv}}$ of NAD in the range $\approx 92$ dyn/cm to $\approx 100$ dyn/cm,  while the MKLV
expression predicts it in the range $\approx 95$ dyn/cm to $\approx 105$ dyn/cm. The predictions of both
expressions for $W_*$ become identical for the case of congruent crystallization; this was also demonstrated
by applying our method for determining   $\sigma^{\mbox{\tiny sv}}$  to the nucleation of nitric acid
trihydrate (NAT) crystals in PSC droplets of stoichiometric composition. 

}
\renewcommand{\baselinestretch}{1} 
\newpage 
\tracinggroups=1
\section{Introduction}

At present, there remains no doubt that type 1 polar stratospheric clouds (PSCs) play a crucial role  in the
atmospheric ozone depletion in general and formation of the so-called ``polar ozone holes"  (dramatic
springtime decrease in stratospheric ozone over polar regions) in particular.$^{1,2}$   Chemical reactions on
the surface of type 1 PSC particles in the wintertime polar  stratosphere trigger$^{3}$ a large, albeit
localized  and seasonal, increase in the amount of chlorine in its active, ozone-destroying radical  forms.  
Under the springtime sunlight, such radicals efficiently destroy ozone molecules in a series of  chain
reactions.  Type 1 PSC formation is also harmful because, through sedimentation,  it removes gaseous nitric
acid  from the stratosphere which would otherwise combine with C$l$O to form less reactive forms of chlorine. 

Since the chemical activity of type 1 PSCs strongly depends on phases present in their particles,  it is
important to  understand both the conditions under which they form and their phase evolution.  In
particular, it is important to predict atmospheric conditions (temperatures, partial pressures, etc...)
under which  liquid droplets of type 1b PSCs freeze, thus transforming into solid particles of type 1a
PSCs. A (parameterized) theoretical description of the crystallization of atmospheric droplets  constitutes
an integral part of  the aerosol/cloud component of regional and global climate models; the improvement of
the former will hence  lead to the improvement of the latter and enhancement of their predictive powers. 

One of the most important characteristics of the crystallization of liquids 
is the crystal nucleation rate.  Until about 2002, crystallization in homogeneous liquids had been
assumed to initiate within the volume of the supercooled phase.$^{4,5}$  Under such an assumption, the rate
of crystallization of a droplet is proportional to its volume.$^{4-6}$   Not surprisingly, until about a
decade ago, experimental data on the crystallization of aqueous nitric acid droplets  had been analyzed
under the assumption that the NAD and NAT nucleation there took place in the interior of droplets. 

However,  Tabazadeh {\em et al.}$^{7}$ re-examined some laboratory data on the homogeneous freezing of
aqueous nitric acid droplets and suggested crystal nucleation to occur ``pseudoheterogeneously"  at the
air-droplet interface. Similar conclusions were drawn for the freezing of pure water droplets.$^{8}$  

Moreover, using  the classical nucleation theory (CNT),
Djikaev {\em et al.}$^{9,10}$ developed a thermodynamic theory of surface-stimulated nucleation. The 
theory  prescribes the condition  under which the  surface of a droplet can stimulate crystal
nucleation therein so  that  the formation of a crystal nucleus with one of its facets at the droplet surface
(``surface-stimulated" mode) is thermodynamically favored over its formation with all the facets {\em within}
the liquid phase (``volume-based" mode).   For both  unary and multicomponent droplets, this condition
coincides with the  condition for the partial wettability of at least one of the crystal nucleus facets by the
liquid.$^{11}$
This effect was experimentally observed for  several systems,$^{12-14}$ including water-ice$^{15}$ 
at temperatures at or  below 0$^{o}$C.  

Clearly, under otherwise identical thermodynamic conditions, the free energy barrier of heterogeneous
nucleation is much lower than that of homogeneous  nucleation.$^{6,11,16,17}$  However, as previously 
shown,$^{18}$ surface-stimulated crystal nucleation  should {\em not} be considered  as a
particular case of  heterogeneous nucleation and its kinetics {\em cannot} be treated by using the  formalism
of heterogeneous nucleation on foreign surfaces (such a conclusion also transpired from some experimental
results on the freezing of both pure water$^{19}$ and aqueous nitric acid$^{20}$ droplets); 
it is much more likely to be a special case of homogeneous crystal nucleation,$^{18}$ 
where one of the facets of a crystal nucleus represents a part of the surface of the (freezing) droplet.

The free energy of formation of such a ``surface-stimulated" crystal nucleus is related to the free energy of
formation of a ``volume-based" nucleus (with all its facets {\em within} the droplet) through an expression 
involving the morphology parameters of the crystal nucleus and its interfacial tensions with both liquid and
vapor phases.$^{18}$ Using that relationship, one can 
determine the solid-vapor interfacial tension via experiments on crystal nucleation in droplets of 
different sizes. 
The  method exploits
the dominance of the surface-stimulated mode of crystal nucleation in small droplets  (of radii $R\lesssim5$
$\mu$m) and its negligibility in large ones (of radii $R\gtrsim20$ $\mu$m). Experimentally
measuring the crystal nucleation rates in such large and small (but otherwise identical) droplets, one can
determine the solid-vapor interfacial tension of that crystal facet  which forms at the droplet surface in a
surface-stimulated mode. 

The proposed method assumes that there exists an analytical expression for the free energy of formation of a 
crystal nucleus {\em within} the liquid phase (i.e., in the volume-based mode). In the case where
crystallization occurs in unary liquids or if a multicomponent crystalline phase forms out of its own, 
stoichiometric melt (solution),  such an expression$^{}$ has been well established (in the framework of CNT) 
and widely used for extracting the liquid-solid interfacial tension from
nucleation experiments.$^{6,21-23}$

For the case where a multicomponent crystalline phase forms out of a non-stoichiometric solution
(incongruent solidification), an analogous 
expression (relating the free energy of formation of a crystal nucleus to its solid-liquid interfacial
tension) exists as well, and it is usually cited$^{21,23-25}$ as due to MacKenzie,  Kulmala, Laaksonen, and 
Vesala$^{26}$ (hereinafter referred to as MKLV) and Pruppacher and Klett.$^{6}$ While we were not able to
locate that particular expression in  the book by Pruppacher and Klett,$^{6}$ MKLV provide that expression
(equation (1) in ref.26) stating that its derivation ``{\em ...follows Pruppacher and Klett [1978, p.
136ff, and p.152], and assumes that all solid phases are pure.}"  Despite the latter assumption, the  MKLV
expression has been widely applied to crystal nucleation in aqueous solutions even when the solid phase was
{\em not} pure ice; in particular, it was used to extract estimates for the crystal-liquid  interfacial
tension of NAD and NAT from the experiments on the freezing of droplets of aqueous nitric
acid.$^{21,23,25,26}$ 

Avoiding such an assumption (that the solid, crystallizing phase is a pure, single-component one),  in this
paper we will present the derivation of an expression for the free energy of formation of a binary crystal
nucleus within a non-stoichiometric solution (incongruent crystallization), thus obtaining an expression,
analogous to that of MKLV, but which more adequately takes account of the presence both components in the
system, including the vapor-gas medium (wherewith the liquid phase is assumed to be in equilibrium).  Applying
our expression to experiments on the freezing of large aqueous nitric acid droplets,$^{25}$ we will then
evaluate the solid-liquid interfacial tensions of NAD crystals in  non-stoichiometric solutions and compare
our estimates with those, obtained on the basis of the MKLV expression.$^{26}$ Furthermore, using these
estimates and applying our previously proposed method (mentioned above) to experiments on the freezing of
small aqueous nitric acid droplets of non-stoichiometric composition,$^{20}$ we will also evaluate the
solid-vapor interfacial tensions of NAD crystals and compare the predictions based on the MKLV expression (for
the free energy of formation of a crystal nucleus)  with those based on our expression. As a further
illustration of that method, we will also present evaluations of the vapor-solid interfacial tensions of NAT
crystals obtained by analyzing the experimental data on the NAT crystallization in large$^{25}$  and
small$^{27}$ aqueous nitric acid droplets of stoichiometric composition.

\section{Free energy of formation of a binary crystal nucleus {\em within} a binary solution of arbitrary
composition}

It should be emphasized that we are interested in only one of several possible pathways whereby liquid
aqueous  NA droplets may transform into solid particles of NA hydrates. Specifically,  we consider the case
where solid (pure) NAD or NAT particles form via {\em  homogeneous} crystallization of aqueous NA droplets.
Thus, (critical) crystal nuclei of NAD/NAT form {\em homogeneously}, without even ice embryos within a
droplet. The possible presence of ice clusters in droplets of very dilute aqueous NA solution is likely to
result in the dominance of the heterogeneous mechanism of binary nucleation in the  freezing of NAD/NAT
particles or even in the nucleation of ice and  subsequent freezing of binary solutions which would lead to
the phase separation and the formation of two-phase composite droplets, partly consisting of pure ice and
partly of solid NAD/NAT (see Bogdan and Molina$^{28}$).

In the framework of the classical thermodynamics, an expression for  the reversible work of homogeneous
formation of a binary crystal nucleus {\em within} a liquid  binary solution can be derived following the
procedure, used in ref.10 to study the effects of adsorption and  dissociation on the thermodynamics of
surface-stimulated crystal nucleation in multicomponent droplets.  In that procedure, the liquid solution
(whether bulk or in droplets) is assumed to   remain in {\em equilibrium} with the corresponding vapor
mixture at all times. 

If the chemical potentials of all the components in the vapor phase are fixed, so are those in the liquid
solution. Under such conditions, the composition of the liquid solution remains constant during crystal
nucleation. Often, however, a liquid solution freeze into a solid of fixed composition, that is  independent
of a variable solution composition. (For example,  nitric acid hydrates can crystallize from an aqueous nitric
acid solution of variable composition.) If the composition of the liquid solution differs from that of the
solid phase, material must be exchanged  between the vapor mixture and the liquid solution to maintain 
equilibrium. This exchange of material can   lead to a change in the vapor and liquid volumes  (note that the
total volume of the system is fixed).  Thus, {\em unlike} the case of crystallization in  single-component
liquids,$^{9}$ it is now necessary to consider the vapor mixture as a part of our system$^{10}$ in order to take
into account the possibility of  formation of a solid phase of fixed composition from a multicomponent
solution of  arbitrary (non-stoichiometric) composition. 

It should be noted, however, that we are interested  only in the nucleation stage 
of the crystallization of
droplets (without considering the stage of crystal growth), which ends much earlier than any phase
separation  effects (such as those studied, e.g., by Bogdan et al.$^{28}$ and Bogdan and Molina$^{29}$)
take place. Indeed,  at this (nucleation) stage, the sub-critical and near-critical crystal clusters are so
small compared to droplet sizes, that the formation of a crystal nucleus  has virtually no effect on the
composition of even relatively very small droplets of type 1b PSCs.    For example, consider a droplet of
radius 10 nm with the mole fraction of NA $\chi=0.3$; such a droplet will  contain about 38511 molecules
total (11553 molecules of NA and 26958 molecules of water). If there forms a crystal nucleus of NAD (with
$\chi^{\delta}=(1/3)\ne\chi$)  consisting of 30 molecules of water and 60 molecules of NA  (typical sizes
of crystal nuclei, according to refs.6, 11, 16) in such a droplet, the composition of its liquid part  will
become  $\chi\approx 0.299922$, changing by less than 0.03\%, i.e., remaining virtually constant.
Therefore, it is reasonable to assume that at the crystal nucleation stage the composition  of the liquid
droplet remains virtually constant even if its composition is markedly non-stoichiometric; hence,   no
phase separation can occur upon the formation of a single crystal nucleus.

Thus, for convenience, the work of formation can be calculated  in the constant $N,\;V,\;T$  ensemble,
where $N$ is  the total number of molecules in the system, $V$ is its volume, and  $T$ is the temperature
(assumed to be uniform  in the system, including the droplet).  Strictly speaking, in this ensemble it is
not possible to fix the chemical potentials of the vapor mixture, since its volume and number of molecules
can vary during crystallization. However, as clear from the foregoing, the variations of the  chemical
potentials that might arise at the nucleation stage during incongruent crystallization, will be rather
negligible.  Likewise, the change in the volume of the liquid droplet due to the formation of a tiny
crystal nucleus therein  can be also neglected owing to small differences in the densities of liquid and
solid phases (and smallness of the  nucleus compared to the droplet). Moreover, in the thermodynamic
limit,  since the crystalline nucleus is usually very small  compared to the bulk liquid solution, the
relative inaccuracy in the free energy due to such variations tends to zero, and the reversible work of
formation of a crystal particle does not depend on the choice of the ensemble (for more detailed discussion
of this issue see, e.g., refs.9 and 10). The equivalence of different thermodynamic potentials (Helmholtz
free energy, Gibbs free energy, and grand thermodynamic potential) in calculating the free energy of
formation of a nucleus of a new phase was first proven by Lee et al.$^{30}$ (see section 2 therein). 

Consider a liquid binary solution (either in a container or in a droplet)  with a binary crystal particle
formed therewithin.  The crystal is of arbitrary shape with $\lambda$ facets. We will use the Greek
superscripts $\alpha$, $\beta$, and $\delta$ to  mark quantities for the liquid, vapor, and crystal, 
respectively, while double Greek superscripts mark quantities at the corresponding interfaces.

\par Assuming the crystallization process to be isothermal  (with the constant uniform temperature $T$
below the temperature $T_m$ of solid-liquid phase equilibrium at given partial pressures)$^{9,10}$ and
taking into  account the foregoing arguments, the reversible work of crystal nucleus formation,  $W$, can
be determined via the difference between $F_{fin}$, the Helmholtz free energy of the system in its final
state (vapor+liquid+crystal), and $F_{in}$, the Helmholtz free energy in the initial state (vapor+liquid):
\beq W=F_{fin}-F_{in}.\eeq
The negligibility of non-isothermal effects during crystal nucleation in droplets is a reasonable assumption  
for both laboratory and atmospheric conditions.$^{6, 11, 16}$ 

Using the formalism of CNT  (based on the capillarity approximation in the framework of the classical
thermodynamics)  and taking into account both dissociation of solute molecules in the solution and  surface
adsorption effects, one can  obtain$^{10}$ the following expression for the reversible work of homogeneous
formation of a  binary crystal particle of a fixed composition {\em within}  a non-stoichiometric liquid
binary solution of different (arbitrary) composition  (incongruent solidification):
\be W&=&-\sum_i\nu^{}_{i}\left(\De H_{i}\ln\Theta-
k_BTz_i^{\alpha}\ln\frac{a_i^{\delta}(\chi^{\delta})}{
a_i^{\alpha}(\chi^{\alpha})}\right)
+\sum_i \sum_{j=1}^{\la} N_{ij}^{\alpha\delta}
(z_{ij}^{\alpha\delta}\mu_{ij}^{\alpha\delta}(\Gamma_j^{\alpha\delta},T)
-z_i^{\alpha}\mu_i^{\alpha}(P^{\alpha},\chi^{\alpha},T))\nonumber\\
&&+\sum_{j=1}^{\lambda}
\sigma^{\alpha\delta}_{j}A^{\alpha\delta}_{j}.\ee
Here,   $\nu_{i}$, $\chi_i=\nu_i/(\nu_1+\nu_2)$, and $\mu_i^{}(P^{},\chi^{},T)$ are the number of
molecules, mole fraction, and chemical potential of component $i\;\;(i=1,2)$, respectively (clearly, the
composition of a binary system is uniquely defined by $\chi\equiv \chi_1$);   $P$ is the pressure, and
$\Theta\equiv T/T_m$.   The first term on the RHS of eq.(2) represents the excess Gibbs free energy of the
molecules in the bulk crystal with respect to  their free energy in the liquid state, and is related to the
partial molecular enthalpy of fusion of all components  $\De H_i\equiv\De H_i(\chi^{\delta})<0$ for  the
case in which the solid and liquid phases have the same composition (congruent  melting/solidification).
The numbers of ions into which a molecule of component $i$ dissociates in the solution and (when adsorbed)
at the facet $j$ are denoted by  $z_i^{\alpha}$ and $z_{ij}^{\alpha\delta}$, respectively.  For components
that   dissociate in the solution or at the interface, the corresponding  chemical potentials, $\mu$'s, and
activities, $a$'s, should be understood as mean ionic chemical potentials and  mean activities,
respectively.  The second term on the RHS of eq.(2) is due to the adsorption of both components at the
facets of the crystal particle.  The surface area and  surface tension of the facet $j$ are denoted by
$A_{j}$ and $\sigma_{j}$, respectively.
The total
number of molecules, chemical potential, and adsorption of component $i$ at facet $j$ are denoted by
$N_{ij}^{},\;\; \mu_{ij}^{}$, and $\Gamma_{ij}^{}$, respectively.  The composite variable
$\Gamma_j^{\alpha\delta}$ is the pair of adsorptions  \{$\Gamma_{1j}^{\alpha\delta},
\;\Gamma_{2j}^{\alpha\delta}$\} of both components, with 
$\Gamma^{\alpha\delta}_{ij}=N_{ij}^{\alpha\delta}/A^{\alpha\beta}_j\;\;(i=1,2)$.  

Note that eq.(2) takes into account the density difference between crystal and liquid phases. On the other
hand, it assumes that in the temperature
range between $T$ and $T_{m}$ the partial enthalpy of fusion $\De H_{i}$
does not change significantly. If this assumption is not accurate enough, then $\De H_{i}$ must be interpreted
as some average value between temperatures $T$ and $T_0$ (such as, e.g., $\De H_{i}=
(\De H_{i}(T_m)+\De H_{i}(T))/2$, in the simplest approximation).

\par Equation (2) allows the composition of the solid phase to differ
from that of the liquid solution. Thus during crystallization 
the vapor mixture
and liquid solution have to continually adjust the equilibrium between them 
if $\chi^{\alpha}$ differs 
from $\chi^{\delta}$ (as is often the case with freezing of aqueous
nitric acid into nitric acid hydrates). Since we work in the closed 
$NVT$ ensemble, the 
composition of the liquid solution may change as it
freezes, and so may the vapor (and hence the liquid) pressures.
In addition, the density of the liquid may differ from that of the
solid. If it does, the crystal formation will induce a change in the
volume of the vapor phase. The volume change can affect 
the partial pressures in the vapor mixture (and
the pressure in the liquid), 
which in turn can induce a change in the liquid composition.
However, all the above changes can be neglected at the stage of crystal
nucleation where the sizes of crystal particles are still so small

\par Introduce a characteristic linear size (``radius") $r$ and a characteristic
surface tension $\sigma^{\alpha\delta}$ for a crystal surrounded by its melt as
\beq r=\left[\frac3{4\pi}\sum\nu_i^{}v_i^{\delta}\right]^{1/3},\;\;\;
\sigma^{\alpha\delta}=\frac1{4\pi R^2}\sum_{j=1}^{\lambda}
\sigma^{\alpha\delta}_jA^{\alpha\delta}_j, \eeq
where $v_i^{\delta}$ is the partial volume per molecule of component $i$ in the solid phase. 
If the crystal had a spherical shape, it would have a radius $r$ and a surface tension
$\sigma^{\alpha\delta}$.

\par As usual, define a critical crystal (nucleus)
as a crystal that is in unstable equilibrium with the surrounding fluid. The linear size and composition of
such a crystal nucleus can be found by solving simultaneous equations (representing the conditions 
for the extremum of the free energy of formation of a critical crystal, i.e., nucleus)$^{4,6,10,11,16-18}$
\beq \left. \partial W/\partial \nu_i \right|_{r_*,\chi_*}=0\;\;\;\;(i=1,2),\eeq 
(with $\chi_*$ bound to be equal to $\chi^{\delta}$ when the solid phase can have only specific
stoichiometric composition, as in the case of formation of NAD or NAT crystals).
(hereafter the subscript ``*'' indicates quantities corresponding to the nucleus). 

Substituting eq.(2) into eq.(4), 
recalling that $\nu_i,\;N_{ij}^{\alpha\delta}\;\;\;(i=1,2; j=1,...,\lambda)$ 
are all independent variables of state of the crystal
particle, and taking account of definitions (3), one can show that 
the characteristic radius ${r_*}$
of the crystal nucleus is$^{10}$ 
\beq r_*=2\sigma^{\alpha\delta}v_i^{\delta}\left(\De H_i(\chi^{\delta})\ln\Theta-k_BTz_i^{}\ln
\frac{a_i^{\delta}(\chi_i^{\delta})}{a_i^{\alpha}(\chi_i^{\alpha})}\right)^{-1}\;\;\;(i=1,2).\eeq 
Alternatively, one can rewrite this expression in a more symmetric form, 
\beq r_*=2\sigma^{\alpha\delta}\bar{v}^{\delta}\left(\De q(\chi^{\delta})\ln\Theta-
k_BT\Big(
\chi^{\delta}z_1\ln\frac{a_1^{\delta}(\chi^{\delta})}{a_1^{\alpha}(\chi^{\alpha})}+
(1-\chi^{\delta})z_2\ln\frac{a_2^{\delta}(\chi^{\delta})}{a_2^{\alpha}(\chi^{\alpha})}
\Big)
\right)^{-1}, \;\;\; \eeq 
where $\De q=\chi^{\delta}\De H_1+(1-\chi^{\delta})\De H_2$ and
$\bar{v}\equiv\bar{v}^{\delta}=\chi^{\delta}v_1^{\delta}+(1-\chi^{\delta})v_2^{\delta}$.

From this point on, let us assign subscript ``1" to nitric acid (NA) and replace it by ``na", while assigning 
subscript ``2" to water and replacing it by ``w". Since 
the work
of crystal formation $W$ as a function of $r$ attains its maximum at $r=r_*$, one can then obtain 
\beq 
W_{*}=\frac{(16\pi/3)(\sigma^{ls})^3\bar{v}^2}
{\big[\Delta q\ln(T/T_m)+ 
k_BT (
\chi^{\delta}\ln\frac{P^{e}_{na}(\chi)}{P^{e}_{na}(\chi^{\delta})}+
(1-\chi^{\delta})\ln\frac{P^{e}_{w}(\chi)}{P^{e}_{w}(\chi^{\delta})} )\big]^2}, 
\eeq
where $\chi$ and $\chi^{\delta}$ are the mole fractions of solute molecules in the liquid solution and 
solid (crystalline) phase, respectively; 
$P^{eq}_{na}(\chi')$ and $P^{eq}_{w}(\chi')$ are the equilibrium vapor pressures of NA and water,
respectively, at a given temperature $T$ 
over a bulk liquid binary solution of composition $\chi'$ (mole fraction of component 1). 

Deriving eq.(7), it was taken into account that  at melting temperature $T_m$,  the activity coefficient 
of each component is the same in both phases for congruent melting, i.e., 
$a_i^{\alpha}(\chi^{\delta},T_m)=a_i^{\delta}(\chi^{\delta},T_m)\;\;\;(i=1,2)$. Besides, they were assumed 
to be weakly dependent on temperature in the vicinity of $T_m$, i.e.,  $\partial
a_i^{\alpha}(\chi^{\delta},T)/\partial T\ll a_i^{\alpha}(\chi^{\delta},T_m)/|T-T_m|$. 

Taking into account the foregoing and comparing eq.(7) with the MLKV expression for $W_*$, 
\beq 
W_{*}=\frac{(16\pi/3)(\sigma^{ls})^3\bar{v}^2}{\big[\Delta q\ln(T/T_m)+
k_BT\ln\frac{P^{e}_{w}(\chi)}{P^{e}_{w}(\chi^{\delta})}\big]^2}, 
\eeq 
(eq.(1) in ref.26), 
it is clear that the principal difference between the two is the absence of the term 
$k_BT(\chi^{\delta})\ln\frac{P^{e}_{na}(\chi)}{P^{e}_{na}(\chi^{\delta})}$ 
from the denominator on the RHS of eq.(8), representing the ratio of the vapor pressures of nitric 
acid in equilibrium with the actual solution and with the solution of stoichiometric composition (equal to
that of the solid phase). This is not surprising, because, as explicitly stated by MKLV,$^{26}$
their expression, equation (8), assumes that the solid phase was a pure solid (ice). What is surprising is 
that eq.(8) was applied to the formation of NAD/NAT crystals
in aqueous NA droplets of non-stoichiometric composition.$^{21,23,25,26}$ 
It should be noted, that  the original expression for $W_*$ by MacKenzie et al. (ref. 22, eq.(13))  was
more adequate  (except for   some typos therein) for incongruent solidification than their corrected
version,$^{26}$ eq.(8) above.  On the other hand, eq.(8) (i.e., equation (1) in ref.26) would be 
correct if, in the denominator of  its RHS, the enthalpy of freezing had been replaced by the partial 
molar enthalpy of freezing of  water and, in the numerator of its RHS, the average volume per
molecule $\bar{v}$ had been replaced by the partial molecular volume of water molecules $v_w^{\delta}\equiv
v_w(\chi^{\delta})$ in  the solid phase of composition $\chi^{\delta}$. 
(As a side note, 
$\bar{v}$  in the numerator of eq.(8) should be the average molecular 
volume in the solid phase rather than in the liquid; however, the differences in $\bar{v}$ for the solid
and liquid phases can be neglected in most atmospheric applications.)

Since both equations (7) and (8) assume the crystal nucleus to be spherical, they allow one determine  
$\sigma^{ls}$, 
average (over all crystal facets) liquid-solid interfacial tension of the nucleus in a solution of
arbitrary composition if $W_*$ is known. Equation (7)
provides
\beq \sigma^{ls}=2W_*\bar{v}\left[ \Delta q\ln(T/T_m)+
k_BT\Big(\chi^{\delta}
\ln\frac{P^{e}_{na}(\chi)}{P^{e}_{na}(\chi^{\delta})}+ 
(1-\chi^{\delta})\ln\frac{P^{e}_{w}(\chi)}{P^{e}_{w}(\chi^{\delta})}\Big)\right]^{-1}.  \eeq
whereas eq.(8) leads to
\beq \sigma^{ls}=2W_*\bar{v}\left[ \Delta q\ln(T/T_m)+
k_BT \ln\frac{P^{e}_{w}(\chi)}{P^{e}_{w}(\chi^{\delta})} \right]^{-1}.  
\eeq

The free energy of formation of a crystal nucleus can be obtained by matching experimental crystal
nucleation rates $J^{\mbox{\tiny exp}}$ with theoretical predictions $J_v^{\mbox{\tiny }}$  of CNT,  i.e.,
by  solving the equation $J^{\mbox{\tiny exp}}=J_v$ with respect to  $W_*$, with   
$$ J_v= \frac{k_BT}{h}\rho_l\mbox{e}^{-\Delta G_{d}/k_BT}\mbox{e}^{-W_{*}/k_BT}, $$   
where $k_B$ and $h$ are the Boltzmann and Planck constants, $T$ is the absolute temperature, $\rho_l$ is 
the number density of liquid molecules,  where $\Delta G_{d}$ is the activation energy for the diffusion of
one molecular unit from the bulk  liquid to the solid (often called the diffusion-activation energy),  and
$W_{*}$ is the reversible work of formation of a crystal nucleus {\em within} the liquid. Thus, 
\beq W_*=-\Delta G_d-k_BT\ln\Big[\frac{h}{k_BT} 
\frac{J_{\mbox{\tiny }}^{\mbox{\tiny exp}}}{\rho_{\mbox{\tiny liq}}}\Big].
\eeq

\section{Two modes of crystal nucleation in droplets}

Equation $J^{\mbox{\tiny exp}}=J_v$, used above, implies that homogeneous crystal nucleation in droplets is 
exclusively volume-based, i.e., that all crystal nuclei are {\em within}  the liquid.  However, as mentioned
above, it was shown both theoretically$^{9,10,18}$ and via analysis$^{7,8}$ of existing experimental data
that  under some conditions the  surface of a droplet can stimulate crystal nucleation therein so  that  the
formation of a crystal nucleus with one of its facets at the droplet surface (``surface-stimulated" mode) is
thermodynamically favored over its formation with all the facets {\em within} the liquid phase
(``volume-based" mode). For both  unary and multicomponent droplets, this condition coincides with the 
condition for the partial wettability of at least one of the crystal nucleus facets by the liquid.$^{6, 11}$ 

Neglecting the contributions from line tensions,  the criterion for whether crystal nucleation in a 
supercooled droplet will or will not be thermodynamically stimulated by the surface 
it has the form$^{9,10}$ 
\begin{equation} \sigma^{sv}_{\lambda}-\sigma^{lv}<\sigma^{ls}_{\lambda},\end{equation}
where $\sigma^{lv}$ the liquid-vapor surface tension, and $\sigma^{ls}_{\lambda}$ and $\sigma^{sv}_{\lambda}$
are the interfacial tensions of crystal facet $\lambda$ in the liquid and in the vapor, respectively. 

Taking this into account, it was suggested$^{7,8,18}$ that the total crystal nucleation rate $J$ in
experiments on droplet freezing can contain two contributions, one due to the volume-based crystal nucleation
and the other to the surface-stimulated mode.   The surface-stimulated mode of crystallization can become
important when the crystallizing liquid  is in a dispersed state, which is the case with the freezing of
atmospheric droplets  and many experiments.  Indeed, since smaller droplets have a higher surface-to-volume
ratio than larger ones, nucleation rates in the former will be higher than in the latter (or in bulk).  Hence 
it is experimentally easier to observe the crystallization in an ensemble of small aerosols than of large
ones,  assuming the total liquid volume is the same in both cases. 

Furthermore, it was argued$^{18}$  that the formation of a crystal nucleus with one of its facets at the
droplet surface  cannot {\em start} preferentially {\em at} the surface, because  the latter does not have
any sites which would make the ordering of the surrounding {\em surface}  molecules  thermodynamically more
favorable than the ordering of interior molecules. On the contrary, the surface layer of a crystal remains
disordered far below the melting   temperature due to weaker constraints on the surface-located molecules
which have a reduced number of neighbors and hence have a higher vibrational  amplitude compared to bulk
ones.  This results in  the formation of a thin  disordered layer on the crystal surface at  temperatures
significantly lower than the melting one. This phenomenon, referred to as premelting,  has been well
established both  experimentally and via molecular dynamics simulations  (see ref.31 and references
therein). Moreover, it was experimentally observed$^{32,33}$   that the premelting of the (0001) face of
hexagonal ice occurs at about $200$ K, far below the lowest temperature reported for homogeneous freezing
of atmospheric droplets.  

Thus,  one can conclude that a crystal nucleus with one facet as a droplet-vapor interface {\em cannot} 
{\em begin} its formation (as a subcritical crystal) at the droplet surface.$^{18}$ Still, the surface of
the droplet {\em can} stimulate crystal  nucleation therein (under condition (12)), but  a crystal cluster
has to begin its initial evolution  {\em homogeneously} in a spherical layer adjacent to the droplet
surface (``sub-surface layer").  When this crystal becomes large enough (due to  the fluctuational growth
usual for nucleation), one of its facets hits the droplet surface and at this moment or shortly thereafter
it becomes a nucleus owing to a drastic change in its thermodynamic state.$^{18}$   any  crystalline
cluster, starting its evolution with its center  in the ``surface-stimulated nucleation" layer, has a
potential to become a nucleus (by means of  fluctuations)  once one of its facets, satisfying eq.(1), meets
the droplet surface. Originally proposed in ref.18, this idea was shown to be plausible in later
experiments$^{19}$ and molecular dynamics simulations.$^{34}$

In this model, the total per-droplet rate of crystal nucleation is the sum of the contributions from both
volume-based and surface-stimulated modes, 
$JV_R=J_v^sV_R^s+J_v(V_R-V_R^s)$,  where 
$J_v^s$ is the number of crystal nuclei, forming in a surface-stimulated mode per unit time 
in unit volume of the surface-stimulated nucleation layer (of total volume $V_R^s$), $J_v$ is the rate of volume-based nucleation, and
$V_R$ is the volume of the droplet. In the framework of CNT, 
$$ J_v^s=\frac{k_BT}{h}\rho_l\mbox{e}^{-\Delta G_{d}/k_BT}\mbox{e}^{-\widetilde{W}_{*}/k_BT}, $$
where the density of molecules in the droplet is assumed to be uniform up to the  dividing  surface, and
$\widetilde{W}_{*}$ is the work of formation of a  surface-stimulated crystal nucleus. 
Introducing a variable $u=(\sigma^{sv}_{\lambda}-\sigma^{lv})/\sigma^{ls}_{\lambda}$, for 
the {\em total} rate of crystal nucleation one can obtain$^{18}$  
\beq J=\Big[1+\big(1-(1-(h_{\lambda}/R)u)^3\big)
\Big(\mbox{e}^{-\frac1{2}(u-1)W_{*}/k_BT}-1\Big)\Big] J_v
\eeq
where $h_{\lambda}$ is the height of a pyramid whereof the apex is at the center of the volume-based crystal
nucleus and the basis is the facet $\lambda$, satisfying eq.(12).

Using eq.(18) from ref.10, one can show, that independent of the shape of the crystal nucleus, its
volume can be found as 
\beq 
V_*=2W_*\bar{v}\left[ \Delta q\ln(T/T_m)+
k_BT\Big(\chi^{\delta}
\ln\frac{P^{e}_{na}(\chi)}{P^{e}_{na}(\chi^{\delta})}+ 
(1-\chi^{\delta})\ln\frac{P^{e}_{w}(\chi)}{P^{e}_{w}(\chi^{\delta})}\Big)\right]^{-1}.  
\eeq
Adopting a more realistic assumption (compared to the assumption of sphericity) that 
the crystal nucleus has a shape of a right prism with an arbitrary aspect ratio $\kappa$, and assuming that
the basal facets of the prism are those that satisfy the criterion of partial wettability, eq.(10), the
quantity $h_{\lambda}$ (defined above)  will be equal to the half-height of the prism. Thus, one can obtain
\beq h_{\lambda }=(\kappa^2V_*)^{1/3}/\sqrt{3}.\eeq

For large droplets of radii $R\gtrsim 20$ $\mu$m, the RHS of eq.(13) reduces$^{18}$ to  $J_v$. 
Thus, if crystal nucleation rates $J^{\mbox{\tiny exp}}_{\mbox{\tiny Large}}$ 
are measured in 
experiments on large droplets, one can find $W_{*}$ (from  $J^{\mbox{\tiny exp}}_{\mbox{\tiny 
Large}}=J_v$), and then extract an average value $\sigma^{ls}$ from eq.(9). If, under identical
experimental conditions, crystal nucleation rates $J^{\mbox{\tiny exp}}_{\mbox{\tiny small}}$ 
are measured in experiments on small
droplets (of radius $R\lesssim5$ $\mu$m), then the rates 
$J^{\mbox{\tiny exp}}_{\mbox{\tiny small}}$ and $J^{\mbox{\tiny exp}}_{\mbox{\tiny Large}}$ 
will be related as 
\begin{equation} \frac{J^{\mbox{\tiny exp}}_{\mbox{\tiny small}}}{J^{\mbox{\tiny exp}}_{\mbox{\tiny
Large}}}= \Big[1+\big(1-(1-(h_{\lambda}/R)u)^3\big)
\Big(\mbox{e}^{-\frac1{2}(u-1)W_{*}/k_BT}-1\Big)\Big].\end{equation} 

Since $W_{*}$ can be extracted from experiments on
freezing of large droplets, so can $\sigma^{ls}$ and $h_{\lambda}$. Therefore,   
equation (16) can be solved with respect to $u$. Assuming that 
$\sigma^{lv}$ is known and that $\sigma^{ls}_{\lambda}\approx\sigma^{ls}$,  
one can determine the solid-vapor interfacial tension $\sigma_{\lambda }^{sv}$ of the
facet $\lambda$ as 
\beq \sigma_{\lambda }^{sv}=\sigma^{lv}+u_{\mbox{\tiny 0}}\sigma_{\lambda }^{ls},\eeq
where $u_{\mbox{\tiny 0}}$ is the (physically meaningful) solution of equation (16). 

Clearly, this method for determining the solid-vapor interfacial tension 
$\sigma_{\lambda}^{sv}$ from experiments on the freezing of aqueous NA droplets, requires the 
experimental data in 
for crystal nucleation rates in large and small droplets,  $J^{\mbox{\tiny exp}}_{\mbox{\tiny Large}}$
and $J^{\mbox{\tiny exp}}_{\mbox{\tiny small}}$, to be for identical experimental conditions, droplets
differing only in their size. Otherwise, its application  provide only rough estimates for 
$\sigma_{\lambda }^{sv}$.

The RHS of eq.(16) is a non-monotonic function of $u$, $f(u)$ (Figure 1),   hence eq.(16) can formally have
two roots: one, $u'$,  with $df(u)/du|_{u'} > 0$, and the other, $u_0$,  with $df(u)/du|_{u_0} < 0$.
However, the root $u'$ is not physically meaningful. Indeed, at constant  $\sigma_{lv}$ and
$\sigma_{\lambda }^{ls}$, one can expect $J^{\mbox{\tiny exp}}_{\mbox{\tiny small}}$ to  decrease with 
increasing $\sigma_{\lambda }^{sv}$ hence, $df(u)/du$ must be negative. 

Note that, while  $\sigma^{lv}$, $\sigma^{ls}$, and $\sigma^{ls}_{\lambda}$ depend on the composition of
the solution in droplets, $\sigma_{\lambda}^{sv}$ can be expected to be independent of whether NAD/NAT 
crystals form in small droplets of aqueous nitric acid of stoichiometric or non-stoichiometric composition,
assuming that, in both cases, the same crystal facet $\lambda$ forms as a part of the droplet-air
interface.

\section{Numerical Evaluations}
In applications to the crystallization of aqueous nitric acid droplets,  the new expression  more
adequately takes  account of the effects of nitric acid vapors compared to the conventional, widely used
expression of MacKenzie, Kulmala, Laaksonen, and Vesala (MKLV).$^{26}$  The predictions of both our
modified expression and the MKLV one for the average  liquid-solid interfacial tension
$\sigma^{\mbox{\tiny ls}}$ of nitric acid dihydrate (NAD) crystals are then compared by applying them to
the analysis of existing experimental data on the incongruent crystallization of aqueous nitric acid
droplets of composition relevant to polar stratospheric clouds. It is shown that predictions for 
$\sigma^{\mbox{\tiny ls}}$ based on  the MKLV expression are usually higher by about 5\% compared to
predictions based on our modified expression. These differences are then transferred into the predictions
of both expressions for the solid-vapor interfacial tensions of NAD crystal nuclei. The latter can be
obtained by using  the method, that we previously developed for determining the solid-vapor interfacial
tension  $\sigma^{\mbox{\tiny sv}}$  of nitric acid dihydrate/trihydrate (NAD/NAT)  crystals using
experiments  on crystal nucleation in aqueous nitric acid (NA) droplets; it exploits the dominance of the
surface-stimulated mode of crystal nucleation in small droplets  (of radii $R\lesssim5$ $\mu$m) and its
negligibility in large ones (of radii $R\gtrsim20$ $\mu$m). Applying our method to existing experimental
data,  our expression for the free energy of formation provides an  estimate for $\sigma^{\mbox{\tiny sv}}$
of NAD to be about $90$ dyn/cm, while the MKLV expression predicts it to be about $95$ dyn/cm. The
predictions of both expressions for $W_*$ become identical for the case of congruent crystallization; this
was also demonstrated by applying our method to the nucleation of nitric acid trihydrate (NAT) crystals in
droplets of stoichiometric composition. 

To numerically illustrate differences between 
the two expressions for $W_*$, MKLV one (eq.(8)) and modified one (eq.(7)), 
one can compare the corresponding estimates for the liquid-solid and vapor-solid interfacial tensions that
can be extracted from the experiments on the freezing of aqueous nitric acid droplets by using these
expressions. An MKLV estimate 
$\sigma^{ls}_{\mbox{\tiny MKLV}}$ for $\sigma_{ls}$ is provided by eq.(10), whereas a modified estimate
$\sigma^{ls}_{\mbox{\tiny DR}}$ is given by eq.(9); using them in eq.(17), one can obtain the corresponding
estimates $\sigma^{sv}_{\mbox{\tiny MKLV}}$ and $\sigma^{sv}_{\mbox{\tiny DR}}$ for $\sigma^{sv}$. 

Estimates for $\sigma_{ls}$ were obtained from the experimental data of Salcedo et al.$^{25}$ for the NAD
and NAT crystal nucleation rates in large aqueous nitric acid droplets (of characteristic radii about $25$
$\mu$m  and various compositions) at stratospherically relevant temperatures. For temperatures and droplet
compositions, at which experiments were not carried out, we used the parameterized expression for the
crystal nucleation rates of NAD and NAT as functions of $T$ and wt\%HNO$_3$ in droplets. 

For the dependence of $P^{eq}_{na}$ and  $P^{eq}_{w}$ on the temperature and composition of aqueous nitric
acid solution we used analytical expressions proposed by   {\em Luo et al.}$^{35}$ to fit their own
experimental results.  The dependence of density $\rho_{liq}$ of nitrate ions in the solution on the
temperature and composition of the solution was taken from ref.36;  for the stoichiometric solution at $193$ K
$\rho_{liq}$ was estimated to be about  $9.2\times 10^{21}$ cm$^{-3}$, and $\bar{v}$ to be about $65.5$
cm$^3$/mol, both based$^{13}$ on a solution density of $1.52$ g/cm$^3$. The latent heat of crystallization was
taken to be$^{21,37}$ $\Delta H\approx -4.8$ kcal/mol.  The temperature and composition  dependence of the
liquid-vapor surface tension  $\sigma^{lv}$  and the temperature dependence of $\Delta G_{d}$ were taken from
refs.38 and 24, respectively.  

Figures 2 and 3 present the $T$- and $\chi$-dependence, respectively, of the effecitve liquid-solid
interfacial tension $\sigma_{\lambda}^{ls}$ of NAD crystal nuclei obtained by using the parameterized
dependence of $J^{\mbox{\tiny exp}}_{\mbox{\tiny Large}}$ on $T$ and  wt\%HNO$_3$, reported in  ref.25 for
large aqueous NA droplets.   In Fig.2, $\sigma_{\lambda}^{ls}$ is shown as a function of temperature $T$ at
the mole fraction of HNO$_3$ in the droplet a) $\chi=0.20$ and b)  $\chi=0.30$,  whereas in Fig.3
$\sigma_{\lambda}^{ls}$ is plotted as a function of $\chi$, at a) $T=190$ K  and b) $T=197$ K.  In both
Figures 2 and 3, the solid and dashed curves correspond to  $\sigma^{ls}_{\mbox{\tiny DR}}$ and
$\sigma^{ls}_{\mbox{\tiny MKLV}}$, respectively. 

As seen in Fig.2, at lower $\chi$'s, in the temperature range from 188 K to 198 K  both
$\sigma_{sv}^{\mbox{\tiny DR}}$ and $\sigma_{sv}^{\mbox{\tiny MKLV}}$ monotonically decrease with increasing
temperature, whereas for higher $\chi$'s they both first increase with increasing $T$, attain a maximum at
about 192-193 K, then decrease.  On the other hand (as seen in both Figs.2 and 3), the difference between them
increases with increasing temperature. However, as expected, this difference becomes zero when the composition
of the liquid solution reaches the stoichiometric composition of the solid phase NAD of $\chi^{\delta}\approx
0.333...$ (compositions $\chi>\chi^{\delta}$ are not shown in Fig.3 because they are irrelevant to
aqueous NA droplets in PSCs under normal stratospheric conditions). One can also notice (Fig.3), that for
$\chi<\chi^{\delta}$, $\sigma^{ls}_{\mbox{\tiny DR}}$ is an increasing  function of $\chi$, while 
$\sigma^{ls}_{\mbox{\tiny MKLV}}$ is a non-monotonic function of $\chi$ (first increasing,  then decreasing
with increasing $\chi$). 

So far, the morphology of NAD and NAT crystal nuclei has not investigated thoroughly, although some
interesting studies have been reported.  The in-situ Fourier transform infrared (FTIR) extinction
spectra of airborne $\alpha$-NAD (low-temperature modification of NAD) microparticles, formed via homogeneous
nucleation in supercooled HNO$_3$/H$_2$O solution  droplets,$^{39}$ suggested 
that the observed $NAD$ crystals are highly  aspherical; modeling them as prolate or oblate spheroids, 
the predominance of oblate shapes with the aspect ratio (defined as the  ratio of the non-rotational and 
rotational axes of the ellipse) in the range from 6 to 20 was reported.$^{39}$ 
The morphology of NAT  crystals forming in aqueous nitric acid solutions of different compositions was
studied$^{40}$ by means of   time-dependent X-ray powder diffraction;   the shape and habit of NAT crystals
were found to vary from platelet-like ones to needle-like ones,  depending on the temperature and composition
of the solution. To take this uncertainty into account, we assumed that NAD crystal nuclei have a shape of a
right prism ($\lambda$ assumed to be the basal facet of the prism) and  evaluated $\sigma_{\lambda}^{sv}$ for
various values of aspect ratio $\kappa$ ($=1, 10, 20, 30, 40, 50$)  in eq.(15).  Our numerical evaluations of
$\sigma_{\lambda}^{sv}$ are presented in Figures 4 and 5.

Figure 4 shows the $\kappa$-dependence  of the solid-vapor interfacial tension $\sigma_{\lambda}^{sv}$ of
facet $\lambda$ of NAD crystal nuclei  obtained by using the experimental data from ref.25  (large aqueous NA
droplets of non-stoichiometric composition) and ref.20  (small aqueous NA droplets of non-stoichiometric
composition) for different temperatures.  The results in Fig.4a were obtained  by using eq.(9) for
$\sigma_{ls}$  (corresponding to the modified expression for $W_*$, eq.(7)), whereas those in Fig.4b were
obtained by using eq.(10) for $\sigma_{ls}$ (corresponding to the MKLV expression for $W_*$, eq.(8)). In both
Figures 4a and 4b, circles represent data due to the experiments on the freezing of aqueous NA droplets of
mean  diameter 225 nm and NA mole fraction 0.27 at temperature 195.8 K;  squares are for droplets of mean
diameter 255 nm and NA mole fraction 0.27 at temperature 192.2 K; diamonds are for droplets of mean diameter
255 nm and NA mole fraction 0.28 at temperature 192.1 K.

Figure 5 presents the $\kappa$-dependence of the solid-vapor interfacial tension $\sigma_{\lambda}^{sv}$  of
facet $\lambda$ of NAT crystals  obtained by using the experimental data of ref.25  (large aqueous NA 
droplets of mean radius 25 $\mu$m and stoichiometric NAT compositions, 54 wt\% HNO$_3$) and  ref.27  
(small aqueous NA  droplets of mean radius 0.38 $\mu$m and stoichiometric NAT composition of 53.7 wt\%
HNO$_3$). The circles correspond to the experiments$^{27}$ at temperature 163.5 K, the squares at 165.5 K, 
and diamonds at 167 K. One can notice that at $T=167$ K, the data points for $\kappa<40$ are absent; this is
due the non-existence of the solution of eq.(16) at such aspect ratios. One can thus conjecture that at
temperature 167 K, crystal nuclei of NAT have highly elongated (needle-like, prolate) shape with the aspect
ratio greater than $40$. 

The thermodynamics and kinetics of crystal nucleation uniquely constrain the size and shape of the critical 
nucleus with a unique, ``native" aspect ratio $\kappa_*$. 
Due to the uncertainties concerning the crystal nucleus morphology, this native aspect ratio $\kappa_*$ is not
known. Hence, it is impossible to accurately determine the value of $\sigma_{\lambda}^{sv}$. 
For all temperatures and solution compositions, as $\kappa$ increases by 5000\% (from 1 to 50), 
$\sigma_{\lambda}^{sv}$ slowly increases by less than about $5$ \% compared to its minimum value at
$\kappa=1$. Most estimates suggest $\sigma_{\lambda}^{sv}$ to
be a weakly increasing function of $T$. Note that extreme caution
should be taken in interpreting these estimates because the experiments on the freezing of large and small
droplets were {\em not} performed under identical thermodynamic conditions (in violation of the fundamental
requirement of eq.(16)). Besides, the scatter of the original experimental rates 
$J^{\mbox{\tiny exp}}_{\mbox{\tiny Large}}$ and $J^{\mbox{\tiny exp}}_{\mbox{\tiny small}}$ would result in
large error bars in estimates of $\sigma_{\lambda}^{sv}$ and cause, for example, the non-monotonic dependence
of $\sigma^{ls}$ on $T$ in Fig.2b.

One can expect that, more accurate information about the native morphology of crystal nuclei 
and more self-consistent  experimental data sets on $J^{\mbox{\tiny exp}}_{\mbox{\tiny
Large}}$ and  $J^{\mbox{\tiny exp}}_{\mbox{\tiny small}}$ will eventually allow one to obtain more accurate 
estimates of $\sigma_{\lambda}^{sv}$ and find its $T$-dependence. 

\section{Concluding Remarks}

Using the formalism of classical thermodynamics in the framework of  the classical nucleation  theory, we
have, with as few assumptions as possible, derived an expression for the reverisble work $W_*$ of formation
of a binary crystal nucleus in a liquid binary solution of non-stoichiometric composition  (incongruent
crystallization). In applications to the crystallization of aqueous nitric acid droplets,  the new
expression  more adequately takes  account of the effects of nitric acid vapors compared to the
conventional, widely used expression of MacKenzie et al.$^{26}$  

Via numerical calculations, we have compared the predictions of the MKLV and modified expressions for the
average liquid-solid interfacial tension $\sigma^{\mbox{\tiny ls}}$ of NAD crystals by applying them to 
the analysis of existing experimental data on the incongruent crystallization of aqueous nitric acid droplets
of composition relevant to PSCs. 

It has been shown that the MKLV-expression-based predictions for $\sigma^{\mbox{\tiny ls}}$  are higher by
about 5\% compared to the predictions obtained by using the modified expression.  Note that 5\% difference
in the interfacial tension $\sigma^{ls}$ is quite significant. Indeed, let us consider the volume-based
crystal nucleation in liquids.  In CNT, the rate of nucleation $J$ is proportional to the exponential of
$W_*$, that is, $J\sim \exp{-W_*/k_BT} $; in turn, $W_*$ is proportional to the third power of the surface
tension $\sigma^{ls}$, i.e., $W_*\sim (\sigma^{ls})^3$, see eqs.(7) and (8).   In order to have a more or
less significant nucleation rate  (i.e., $J>1$  cm$^{-3}$s$^{-1}$,  typical values for $W_*$ are usually of
the order of $30-50$ thermal units $k_BT$ (see, e.g., refs.23,24). Thus, one can easily show, that the
difference of 5\% in the surface tension  $\sigma^{ls}$ would lead to the difference of {\em two orders of
magnitude} in theoretical predictions for $J$. If for example, atmospheric models predicted the crystal
nucleation rate $J$ to be about $10^5$ cm$^{-3}$s$^{-1}$ based on the MKLV expression for $W_*$, then they
would now predict  $J$ to be about $10^7$ cm$^{-3}$s$^{-1}$ based on our modified expression for $W_*$. 

These differences also
transpire  in the predictions of both expressions for the solid-vapor interfacial tension $\sigma^{\mbox{\tiny
sv}}$ of NAD crystal nuclei which can be evaluated from experimental data on  crystal nucleation rates
$J^{\mbox{\tiny exp}}$  in aqueous nitric acid droplets. To extract $\sigma^{\mbox{\tiny sv}}$ from data on 
$J^{\mbox{\tiny exp}}$, it
is necessary to measure the latter both in small droplets (of radii $R\lesssim5$ $\mu$m), wherein  the
surface-stimulated mode of crystal nucleation is dominant, and in large ones  (of radii $R\gtrsim20$ $\mu$m),
wherein the surface-stimulated mode of crystal nucleation is negligible (compared to the volume-based one);
except for the linear size, both small and large droplets must be under  identical thermodynamic conditions. 
The crucial idea of the method is that, even in the  surface-stimulated mode, when a crystal nucleus forms
with one of its facets constituting a part of  the droplet surface, it initially emerges (as a sub-critical
cluster) {\em homogeneously} in the  sub-surface layer, {\em not} ``pseudo-heterogeneously"  at the surface. 

Applying  that method for determining $\sigma^{\mbox{\tiny sv}}$ to existing$^{16,25}$ experimental data,  our
expression for $W_*$ provides an  estimate for $\sigma^{\mbox{\tiny sv}}$ of NAD to be in the range from about
$92$ dyn/cm to about $101$ dyn/cm, while the MKLV expression predicts it to be in the range from 
about $95$ dyn/cm to about $105$ dyn/cm. Both expressions for $W_*$ become
identical for the case of congruent crystallization; this was also demonstrated by applying our method for
determining $\sigma^{\mbox{\tiny sv}}$ to the NAT  crystal nucleation in aqueous nitric acid droplets of
stoichiometric composition. We obtained physically sound and well-behaved estimates of the solid-vapor
interfacial tension of NAT crystals, from about $95$ dyn/cm to about $110$ in the temperature  range from
163.5 K to  about 167 K.

\section*{References}
\begin{list}{}{\labelwidth 0cm \itemindent-\leftmargin} 

\item $^{1}$S. Solomon, Stratospheric ozone depletion: A review of concepts
and history. Rev. Geophys. {\bf 37}, 275-316 (1999).
\item $^{2}$D. Lowe and R. MacKenzie, 
Review of Polar Stratospheric Cloud Microphysics and Chemistry. J. Atmos. Solar-Terrest Phys. 
{\bf 70}, 13-40 (2008).
\item $^{3}$S. Solomon, The mystery of the Antarctic Ozone ``Hole".
Rev. Geophys. {\bf 26}, 131-148 (1988).
\item $^{4}$M. V\"olmer, {\it Kinetik der Phasenbildung} (Teodor
Steinkopff, Dresden und Leipzig, 1939).
\item $^{5}$D. Turnbull and J.C. Fisher, Rate of Nucleation in Condensed Systems. 
J. Chem. Phys. {\bf 17}, 71-73 (1949).
\item $^{6}$H. R. Pruppacher and  J. D. Klett., {\it Microphysics of clouds 
and precipitation}. (Kluwer Academic Publishers, Norwell, 1997).
\item $^{7}$A. Tabazadeh, Y. S. Djikaev, P. Hamill, and H. Reiss, Laboratory evidence for surface nucleation
of solid polar stratospheric cloud particles. J. Phys. Chem. A {\bf 106}, 10238-10246 (2002).
\item $^{8}$A. Tabazadeh, Y. S. Djikaev, and H. Reiss, Surface crystallization of supercooled water in clouds.
Proc. Natl. Acad. Sci. USA {\bf 99}, 15873 (2002).
\item $^{9}$Y. S. Djikaev, A. Tabazadeh, P. Hamill, and H. Reiss, Thermodynamic conditions for the
surface-stimulated crystallization of atmospheric droplets. J.Phys.Chem. A {\bf 106}, 10247 (2002).
\item $^{10}$Y. S. Djikaev, A. Tabazadeh, and H. Reiss, Thermodynamics of crystal nucleation in
multicomponent droplets: Adsorption, dissociation, and surface-stimulated nucleation. 
J.Chem.Phys. {\bf 118}, 6572-6581 (2003).
\item $^{11}$R. Defay, I. Prigogine, A. Bellemans, and D. H. Everett,  
{\it Surface Tension and Adsorption} (John Wiley, New York, 1966).
\item $^{12}$B. Mutaftschiev, and J. Zell, Interfacial energy and cadmium growth in molten baths. 
Surf. Sci. {\bf 12}, 317 (1968).
\item $^{13}$G. Grange, R. Landers, and B. Mutaftshiev, 
Contact angle and surface morphology of KCl crystal-melt interface studied by bubble-method
Surf. Sci. {\bf 54}, 445-462 (1976).
\item $^{14}$D. Chatain and P. Wynblatt, in {\em Dynamics of Crystal Surfaces and Interfaces}, 
Ed. P.M.Duxbury and T.J.Pence, 53-58 (Springer, NY, 2002).
\item $^{15}$M. Elbaum, S. G. Lipson,  and J. G. Dash, Optical study of surface melting on ice.
J. Cryst. Growth {\bf 129}, 491-505 (1993).
\item $^{16}$N. H. Fletcher, {\it The physics of rainclouds}. (University Press, Cambridge, 1962).
\item $^{17}$E. Ruckenstein and G. Berim, 
{\it Kinetic theory of nucleation}. (New York, 2016).
\item $^{18}$Y. S. Djikaev, 
Effect of the Surface-Stimulated Mode on the Kinetics of Homogeneous Crystal Nucleation
in Droplets. J. Phys. Chem. A {\bf 112}, 6592-6600 (2008).
\item $^{19}$T. Kuhn, M. E. Earle, A. F. Khalizov, and J. J. Sloan, Size dependence of volume and surface nucleation
rates for homogeneous freezing of supercooled water droplets. Atmos. Chem. Phys. {\bf 11},
2853-2861 (2011).
\item $^{20}$O. Stetzer, O. M\"ohler, R. Wagner, S. Benz, H. Saathoff, H. Bunz, and O. Indris,  
Homogeneous nucleation rates of nitric acid dihydrate (NAD) at simulated stratospheric conditions - Part I:
Experimental results. Atmos. Chem. Phys. {\bf 6}, 3023-3033 (2006).
\item $^{21}$R .S. Disselkamp, S. E. Anthony, A. J. Prenni, T. B. Onasch, and M. A. Tolbert, 
Crystallization kinetics of nitric acid dihydrate aerosols, J. Phys. Chem. {\bf 100}, 
9127-9137 (1996).
\item $^{22}$A. R. MacKenzie, M. Kulmala, A. Laaksonen, T. Vesala, On the theories of type 1 polar
stratospheric cloud formation. J. Geophys. Res. {\bf 100}, 11275-11288 (1995).
\item $^{23}$A. J. Prenni, T. B. Onasch, R. T. Tisdale, R. L. Siefert, and M. A. Tolbert,
Composition-dependent freezing nucleation rates for HNO3/H20 aerosols resembling gravity-perturbed
stratospheric particles. J. Geophys. Res. {\bf 103}, 28439-28450 (1998).
\item $^{24}$R. T. Tisdale, A. M. Middlebrook, A. J. Prenni, and M. A. Tolbert, 
Crystallization kinetics of HNO3/H20 films representative of polar 
stratospheric clouds. J. Phys. Chem. A {\bf 101}, 2112-2119 (1997).
\item $^{25}$D. Salcedo, L. T. Molina, and M. J. Molina, Homogeneous 
freezing of concentrated aqueous nitric acid solutions at polar
stratospheric temperatures. J. Phys. Chem. A {\bf 105}, 1433-1439 (2001).
\item $^{26}$A. R. MacKenzie, M. Kulmala, A. Laaksonen, and T. Vesala, 
Correction to ``On the theories of type 1 polar stratospheric cloud formation" by A.R.Mackenzie, et al. 
{\bf 102}, 19729-19730 (1997).
\item $^{27}$A. K. Bertram and J. J. Sloan, The nucleation rate constants and freezing 
mechanism of nitric acid trihydrate aerosol under stratosphefic conditions, 
J. Geophys. Res. {\bf 103}, 13,261-13,265 (1998).
\item $^{28}$A. Bogdan, M. J. Molina, H. Tenhu, E. Mayer, T. Loerting, 
Formation of mixed-phase particles during the freezing of polar stratospheric ice clouds. 
Nature Chemistry {\bf 2} (3), 197-201 (2010). 
\item $^{29}$A. Bogdan and M. J. Molina, 
Why Does Large Relative Humidity with Respect to Ice Persist in Cirrus Ice Clouds? 
J. Phys. Chem. A  {\bf 113}, 14123-14130. 
\item $^{30}$D. J. Lee, M. M. Telo da Gama, and K. E. Gubbins, A micrsoscopic theory for spherical
interfaces: Liquid drops in the canonical ensemble. J. Chem. Phys. {\bf 85}, 490-499 (1986). 
\item $^{31}$Y. Djikaev and E. Ruckenstein, A kinetic model for the premelting of a crystalline structure,
Physica A {\bf 387}, 134-144 (2008). 
\item $^{32}$X. Wei, P. B. Miranda, and Y. R. Shen, Surface vibrational spectroscopic study of surface melting of ice. 
Phys. Rev. Lett. {\bf 86}, 1554-1557 (2001). 
\item $^{33}$X. Wei and Y. R. Shen, Vibrational spectroscopy of ice interfaces. Applied Phys. B  
{\bf 74}, 617-620 (2002).
\item $(34)$S. Toxvaerd, N. Larsen, and J. C. Dyre, 
Simulations of crystallization in supercooled nanodroplets in the
presence of a strong exothermic solute: The mystery of the Antarctic ozone ``hole".
J. Phys. Chem. C {\bf  115},  12808-12814 (2011).
\item $^{35}$B. Luo, K. S. Carslaw, T. Peter, and S. L. Clegg, 
Vapour pressures of H2SO4/HNO3/HCl/HBr/H2O solutions to low stratospheric temperatures.
Geophys. Res. Lett. {\bf 22}, 247-250 (1995).
\item $^{36}$E. Martin, C. George, and P. Mirabel, 
Densities and Surface Tensions of H2SO4/HNO3/H20 solutions.  
Geophys. Res. Lett. {\bf 27}, 197-200 (2000).
\item $^{37}$P. J. Wooldridge, R. Zhang, and M. J. Molina, 
Phase equilibria of H2SO4, HNO3, and HCl hydrates and the composition of polar stratospheric clouds.
Geophys. Res. Lett. {\bf 100}, 1389-1396 (1995).
\item $^{38}$V. Granzhan and S. Laktionova,  The  densities,  viscosities,  and
surface tensions of aqueous nitric acid solutions. 
Russian J. Phys. Chem. {\bf 49}, 1448 (1975).
\item $^{39}$R. Wagner, O. M\"ohler, H. Saathoff, O. Stetzer, and U. Schurath, 
Infrared Spectrum of Nitric Acid Dihydrate: Influence of Particle Shape. 
J. Phys. Chem. A {\bf 109}, 2572-2581 (2005).
\item $^{40}$H. Grothe, H. Tizek, D. Waller, and D. J. Stokes, 
The crystallization kinetics and morphology of nitric acid trihydrate.
Phys. Chem. Chem. Phys. {\bf 8}, 2232-2239 (2006).

\end{list}

\newpage 
\subsection*{Captions} 
to  Figures 1 to 5 of the manuscript {\sc
``Free energy of formation of a crystal nucleus in incongruent solidification: Implication for
modeling the crystallization of aqueous nitric acid droplets in polar stratospheric clouds"
}  by {\bf  Yuri S. Djikaev} and {\bf Eli Ruckenstein}. 
\subsubsection*{}
\vspace{-1.0cm}   
Figure 1. Typical dependence of the RHS of eq.(14) on the variable $u$. 
\vspace{0.3cm}\\ 
Figure 2. The $T$-dependence of the effecitve liquid-solid  interfacial tension $\sigma_{\lambda}^{ls}$ of NAD
crystal nuclei obtained by using the parameterized dependence$^{25}$ of $J^{\mbox{\tiny exp}}_{\mbox{\tiny
Large}}$ on $T$ and  wt\%HNO$_3$  for the non-stoichiometric mole fractions of HNO$_3$ in large aqueous NA
droplets:  a) $\chi=0.20$; b)  $\chi=0.30$.   The solid and dashed curves correspond to
$\sigma^{ls}_{\mbox{\tiny DR}}$ (given by eq.(9)) and $\sigma^{ls}_{\mbox{\tiny MKLV}}$ (given by eq.(10)),
respectively. 
\vspace{0.3cm}\\ 
Figure 3. The $\chi$-dependence of the effecitve liquid-solid
interfacial tension $\sigma_{\lambda}^{ls}$ of NAD crystal nuclei obtained by using the parameterized
dependence$^{25}$  of $J^{\mbox{\tiny exp}}_{\mbox{\tiny Large}}$ on $T$ and  wt\%HNO$_3$, 
for large aqueous NA droplets of non-stoichiometric composition at a) $T=190$ K  and b) $T=197$ K. 
The solid and dashed curves correspond to $\sigma^{ls}_{\mbox{\tiny DR}}$ (given by eq.(9)) and
$\sigma^{ls}_{\mbox{\tiny MKLV}}$ (given by eq.(10)), respectively. 
\vspace{0.3cm}\\ 
Figure 4. The $\kappa$-dependence  of the solid-vapor interfacial tension $\sigma_{\lambda}^{sv}$  (dyn/cm)
of facet $\lambda$ of NAD crystal nuclei  due to the experimental data from ref.25  (large aqueous NA
droplets) and ref.20  (small aqueous NA droplets).  The results in Fig.4a were obtained  by using  eqs.(9)
and (17), whereas those in Fig.4b were obtained by using eqs.(10) and (17).  Circles represent experiments
on droplets of  mean  diameter 225 nm and NA mole fraction 0.27 at temperature 195.8 K;  squares are for
droplets of mean diameter 255 nm and NA mole fraction 0.27 at temperature 192.2 K; diamonds are for
droplets of mean diameter 255 nm and NA mole fraction 0.28 at temperature 192.1 K.
\vspace{0.3cm}\\ 
Figure 5. The $\kappa$-dependence of the solid-vapor interfacial tension $\sigma_{\lambda}^{sv}$  (dyn/cm) of
facet $\lambda$ of NAT crystal nuclei due to the experimental data of ref.25  (large aqueous NA 
droplets of mean radius 25 $\mu$m and stoichiometric NAT compositions, 54 wt\% HNO$_3$) and  ref.27  
(small aqueous NA  droplets of mean radius 0.38 $\mu$m and NAT composition of 53.7 wt\% 
HNO$_3$). Circles correspond to the experiments$^{27}$ at temperature 163.5 K, the squares at 165.5 K, 
and diamonds at 167 K.  

\newpage
\begin{figure}[htp]
\begin{center}\vspace{1cm}
\includegraphics[width=8.3cm]{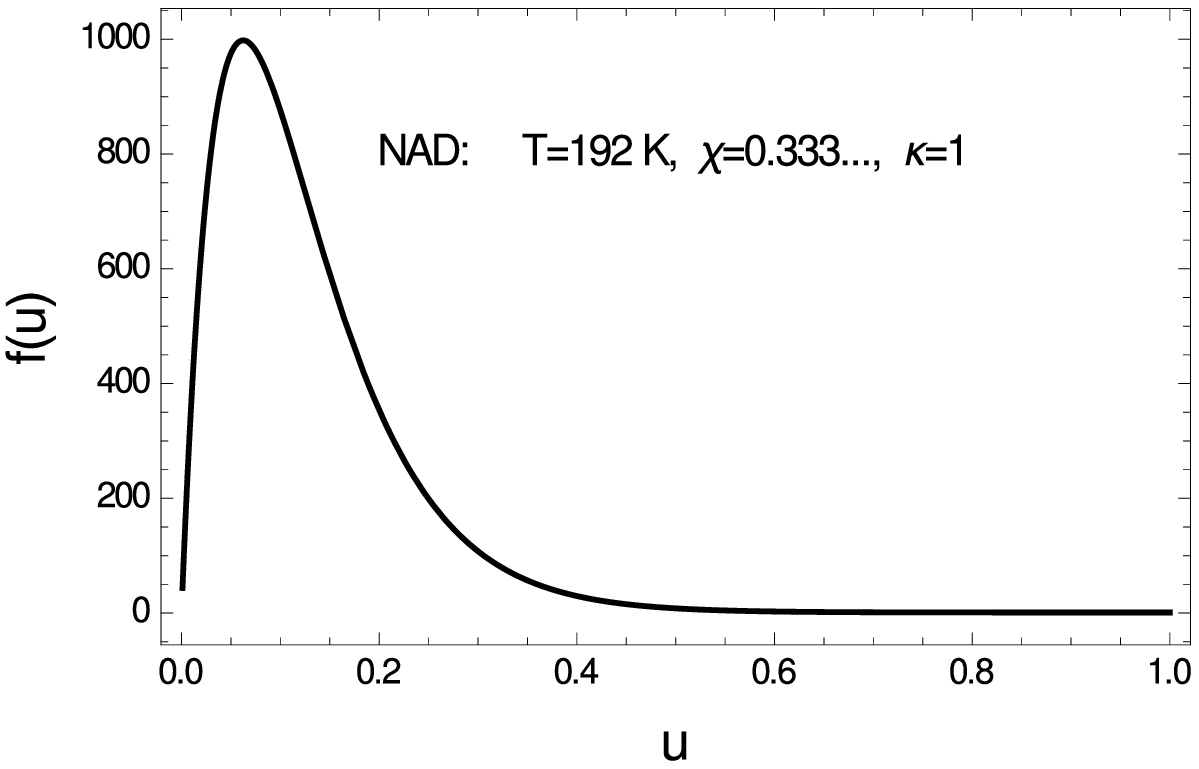}\\ 
\caption{\small }
\end{center}
\end{figure}

\newpage
\begin{figure}[htp]\vspace{-1cm}
	      \begin{center}
$$
\begin{array}{c@{\hspace{0.3cm}}c} 
              \leavevmode
      	      \vspace{3.3cm}
	\leavevmode\hbox{a) \vspace{1cm}} &   
\includegraphics[width=8.3cm]{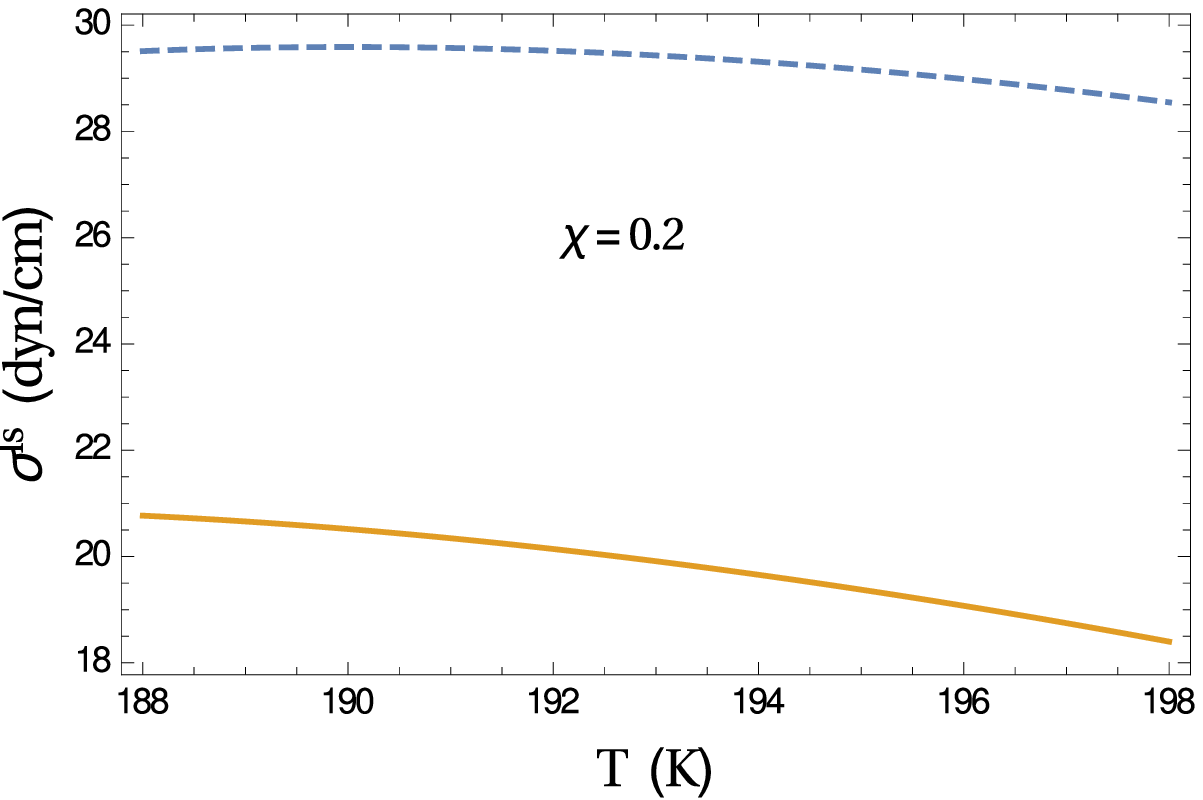}\\ [1.3cm] 
      	      \vspace{1.0cm}
	\leavevmode\hbox{b) \vspace{1cm}} &  
      	      \vspace{0.0cm}
\includegraphics[width=8.3cm]{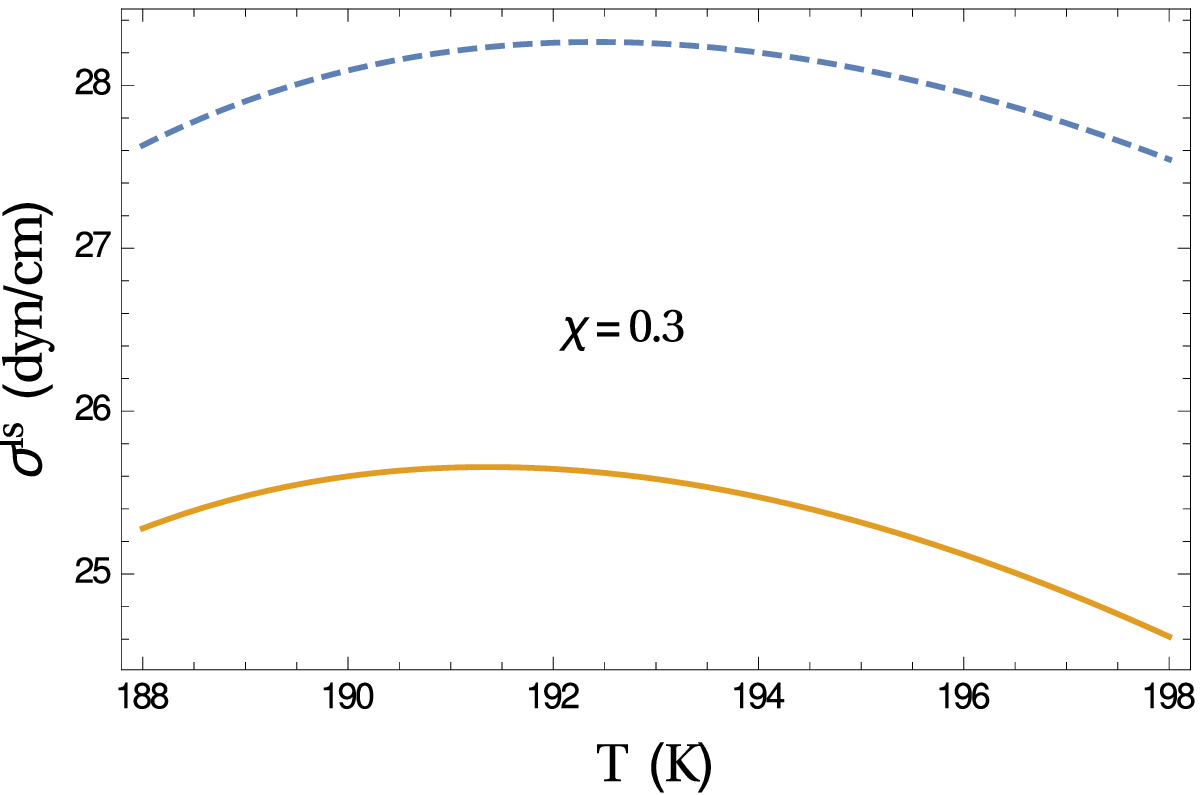}\\ [0.3cm] 
\end{array}  
$$  

	      \end{center} 
            \caption{\small } 
\end{figure}

\newpage
\begin{figure}[htp]\vspace{-1cm}
	      \begin{center}
$$
\begin{array}{c@{\hspace{0.3cm}}c} 
              \leavevmode
      	      \vspace{3.3cm}
	\leavevmode\hbox{a) \vspace{1cm}} &   
\includegraphics[width=8.3cm]{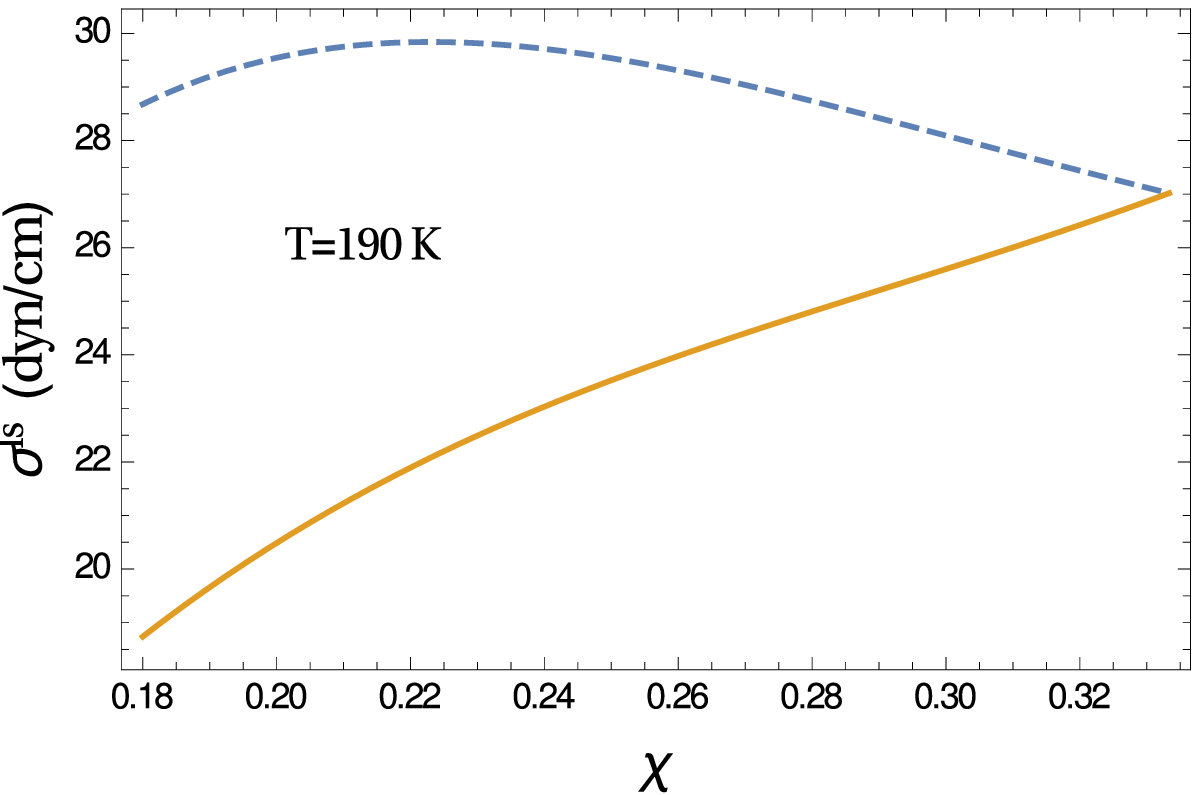}\\ [1.3cm] 
      	      \vspace{1.0cm}
	\leavevmode\hbox{b) \vspace{1cm}} &  
      	      \vspace{0.0cm}
\includegraphics[width=8.3cm]{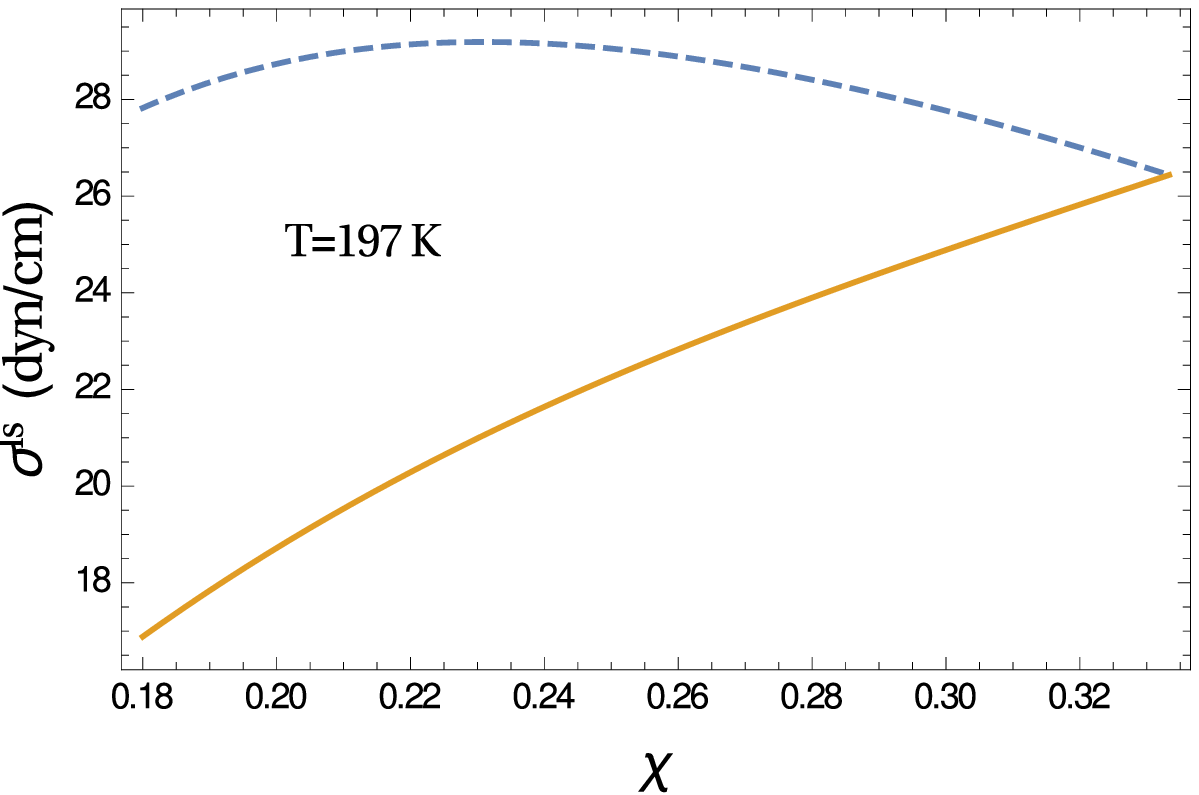}\\ [0.3cm] 
\end{array}  
$$  

	      \end{center} 
            \caption{\small } 
\end{figure}

\newpage
\begin{figure}[htp]\vspace{-1cm}
	      \begin{center}
$$
\begin{array}{c@{\hspace{0.3cm}}c} 
              \leavevmode
      	      \vspace{1.3cm}
	\leavevmode\hbox{a) \vspace{1cm}} &   
\includegraphics[width=8.3cm]{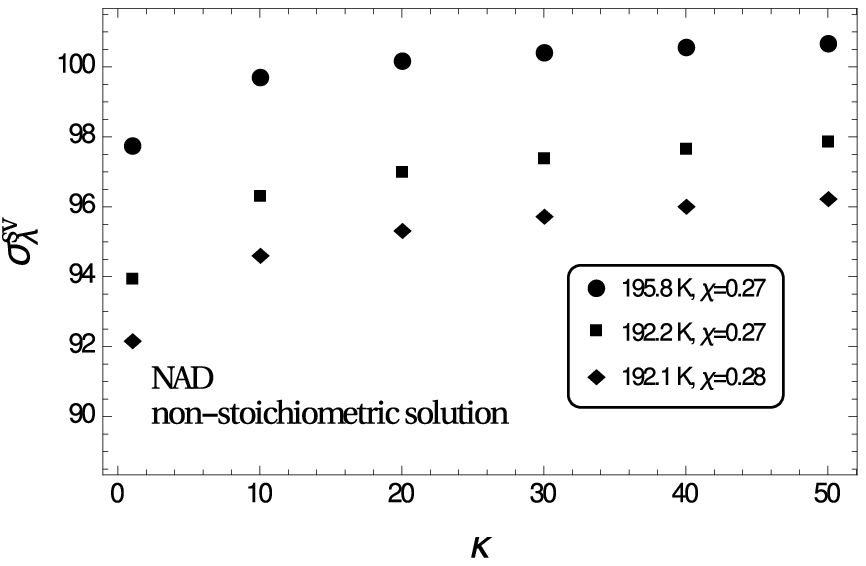}\\ [0.3cm] 
      	      \vspace{1.0cm}
	\leavevmode\hbox{b) \vspace{1cm}} &  
      	      \vspace{0.0cm}
\includegraphics[width=8.3cm]{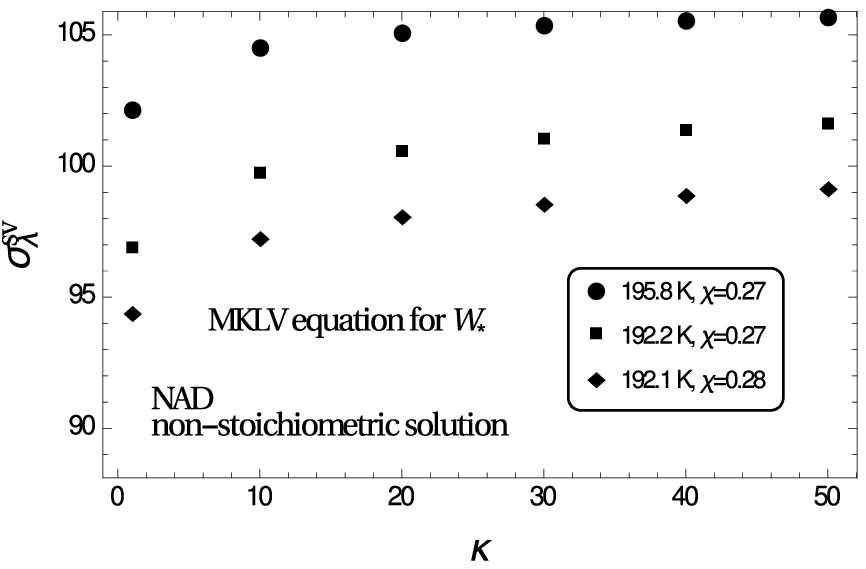}\\ [0.3cm] 
\end{array}  
$$  

	      \end{center} 
            \caption{\small } 
\end{figure}

\newpage
\begin{figure}[htp]
\begin{center}\vspace{1cm}
\includegraphics[width=8.3cm]{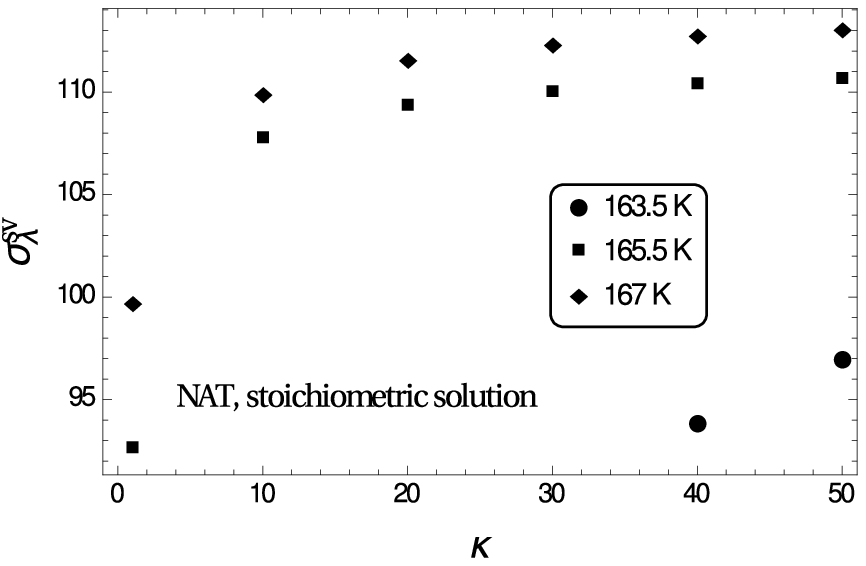}\\ 
\caption{\small }
\end{center}
\end{figure}

\end{document}